\documentclass[12pt]{article}
\pdfoutput=1

\usepackage[pdftex]{graphicx}
\usepackage{lscape} 
\usepackage{array}

\usepackage{fix-cm}

\usepackage[makeroom]{cancel}
\usepackage[normalem]{ulem}
\usepackage{amsmath,amssymb}
\usepackage{slashed}
\usepackage{xcolor}
\usepackage{cite}
\usepackage{pdflscape}
\usepackage{multirow}
\usepackage[thinlines]{easytable}
\usepackage{url}
\usepackage[utf8]{inputenc}
\usepackage{longtable}
\usepackage{booktabs}

\newcommand{\eq}[1]{\begin{equation} #1 \end{equation}}
\newcommand{\eqa}[1]{\begin{eqnarray} #1 \end{eqnarray}}

\newcommand{\Eq}[1]{Eq.~(\ref{#1})}

\newcommand{\App}[1]{Appendix~\ref{#1}}

\newcommand{\Fig}[1]{Figure~\ref{#1}}
\newcommand{\cL}{{\cal L}}

\begin{document}


\begin{titlepage}

\vspace*{-2cm}
\begin{flushright}
\end{flushright}

\vspace{2.2cm}

\begin{center}
\bf
\fontsize{19.6}{24}\selectfont
Specific three-loop contributions to $b\to s\gamma$ associated with the current-current operators
\end{center}

\vspace{0cm}

\begin{center}
\renewcommand{\thefootnote}{\fnsymbol{footnote}}
{Christoph Greub$^a$, Hrachia M. Asatrian$^b$, Francesco Saturnino$^a$ and Christoph Wiegand$^{a}$}
\renewcommand{\thefootnote}{\arabic{footnote}}
\setcounter{footnote}{0}

\vspace{.8cm}
\centerline{${}^a$\it Albert Einstein Center for Fundamental Physics, Institute for Theoretical Physics,}
\centerline{\it University of Bern, CH-3012 Bern, Switzerland}
\vspace*{2.5mm}
\centerline{${}^b$\it Yerevan Physics Institute, 0036 Yerevan, Armenia}
\vspace{2.5mm}

\vspace*{.2cm}

\end{center}

\vspace*{10mm}
\begin{abstract}\noindent\normalsize
We work out a specific class of three-loop diagrams (of order $\alpha_s^2$) contributing
to the decay amplitude for $b \to s \gamma$ associated with the current-current operators
$O_1$ and $O_2$ at the physical value of the charm-quark mass $m_c$.
For many of the considered diagrams we were able to solve the master integrals using differential equations in the canonical form. For some diagrams we did not  find a transformation to canonical form and therefore calculated the corresponding master integral directly as an expansion around $z=m_c^2/m_b^2=0$, retaining power terms up to $z^{10}$ and keeping the accompanying  $\log(z)$
terms to all powers. The results for the sum of all considered diagrams are given in tabular form, while
contributions of individual diagrams (or combinations thereof) are given in electronic form.

\end{abstract}

\end{titlepage}
\newpage 

\renewcommand{\theequation}{\arabic{section}.\arabic{equation}} 

\setcounter{tocdepth}{2}
\tableofcontents


\newpage

\section{Introduction}
\label{sec:intro}
\setcounter{equation}{0}
The (weak) rare decays of  $B$-mesons have been the focus point of theorists and
experimentalists for some time, which is due to the potential they provide
for the tests of the Standard Model (SM) at scales of several hundreds of
GeV. Getting experimental information on rare decays puts strong
constraints on the extensions of the SM, or can lead to disagreements with
the SM predictions, providing evidence for new physics. To make a
rigorous comparison between experiment and theory, one has to get refined
theoretical predictions for the rare decay at hand.
For the inclusive rare $B$-decays, the perturbative strong interaction
effects result in sizable contributions.

In view of the expected increase in precision for the experimental
measurements of the decay $B\to X_s \gamma$, a full next-to-next-to-leading
logarithmic order (NNLL) calculation is necessary to reduce the
theoretical uncertainties and to enable us to perform a rigorous
comparison with existing and future experimental data.
The first estimate of the branching ratio at $O(\alpha_s^2)$, leading to ${\cal
B}(B\to X_s \gamma)=(3.15\pm 0.23)\times 10^{-4}$, was done in
\cite{Misiak:2006zs},  which is consistent with the experimental averages
at the 1.2 $\sigma$ level. An updated version for this branching ratio,
incorporating all results for NNLL contributions  and lower-order
perturbative corrections that have been calculated after 2006,
was published in our paper \cite{Misiak:2015xwa}. For the CP- and
isospin-averaged branching ratio we found
${\cal B}(B\to X_s \gamma)=(3.36\pm 0.23)\times 10^{-4}$
which is in agreement with the current experimental averages.

It is well known that a part of  the $\alpha_s^2$   contributions in
\cite{Misiak:2015xwa} was obtained via interpolation, i.e., using the results
obtained through the large $m_c$ asymptotic expansion on one hand and the results for $m_c=0$ on the other hand.
In the process of evaluating $\alpha_s^2$ corrections directly at the physical
value of $m_c$ in \cite{Misiak:2020vlo}, the part stemming from diagrams with
closed fermion loops  on gluon lines that contribute to the
interference of the current-current and photonic dipole operators was
calculated, extending previous work on such contributions \cite{Bieri:2003ue}.

The present paper is devoted to the computation of virtual $\alpha_s^2$ corrections to the decay amplitude for $b \to s \gamma$ associated with the 
current-current operators\footnote{The present paper is an extension of
our old work on the corresponding  $\alpha_s^1$ corrections \cite{Greub:1996tg}.}
$O_{1}$ and $O_{2}$, where we concentrate on contributions which do {\it not} involve closed fermion loops on gluon lines. 
The corresponding three-loop diagrams are rather complicated to calculate and we therefore divide the complete work into several (gauge invariant)
classes of diagrams. In this paper we describe in detail our computational methods and explicitly work out those diagrams where no gluons are
touching the $b$-quark line (see \Fig{fig:diagsleg}). 

The remainder of this paper is organized as follows: In section 2 we briefly present the theoretical framework and a few conventions. In section 3
we decompose the decay amplitude into form factors which can be written as linear combinations of scalar integrals (SIs); we then reduce the SIs
to master integrals (MIs) and formulate differential equations for the latter. Furthermore we discuss that two methods are needed to work out the master integrals. In section 4 we present the method based on the canonical form of the differential equations which allows us to analytically work out the MIs for many diagrams. However, for four diagrams we did not manage to transform the corresponding MIs into canonical form. Therefore, in section 5,
we present another method to compute these MIs which is based on an expansion in powers and logarithms of $z$ ($z=m_c^2/m_b^2$) around $z=0$. In
section 6 we write the results for all diagrams calculated in this paper as an expansion on this type, retaining terms up to $z^5$ (and keeping
all the accompanying powers of $\log(z)$). In section 7 we summarize our work
and give a short outlook. The results for individual diagrams (or sets
thereof) are submitted in electronic form together with the paper, as described in appendix \ref{app:Bi}.

\section{Theoretical framework}
\label{sec:framework}
\setcounter{equation}{0}

$B$-meson or $b$-quark decay amplitudes are usually calculated within the Weak Effective Theory (WET) where the SM particles with EW-scale masses have been integrated out. The WET lagrangian then contains QCD and QED interactions, and a tower of higher dimensional local operators which is typically truncated at dimension six~\cite{Buchalla:1995vs,Aebischer:2017gaw}. The part of the WET Lagrangian which is relevant for the contributions discussed in this paper is:
\eq{
\cL_{\rm WET} = \cL_{\rm QCD} + \cL_{\rm QED} + \frac{4G_F}{\sqrt2} V_{ts}^*V_{tb} \Big[
C_1 O_1 + C_2 O_2  + C_7 O_7 
\Big]
\label{LWET}
}
where
\begin{align}
&O_1 = (\bar s \gamma_\mu P_L T^a c)(\bar c \gamma^\mu P_L T^a b)\ ,
&&O_2 = (\bar s \gamma_\mu P_L c)(\bar c \gamma^\mu P_L b) \ ,
\nonumber\\[2mm]
&O_7 = \frac{e}{16\pi^2} m_b (\bar s \sigma_{\mu\nu} P_R b) F^{\mu\nu}\ .
&&
\end{align}
We use the following conventions:
$P_{R,L}=(1\pm\gamma_5)/2$,
$\sigma_{\mu\nu}\equiv (i/2) [\gamma_\mu,\gamma_\nu]$,
the covariant derivative is given by $D_\mu q = (\partial_\mu + i e Q_q A_\mu + i g_s T^A G^A_\mu)q$, and $m_b=m_b(\mu)$ denotes the $\overline {\rm MS}$ $b$-quark mass. In our calculation of order $\alpha_s^2$ corrections from $O_{1,2}$, the scheme dependence of $m_b$ is a higher order effect. Furthermore, we will neglect the strange quark mass throughout our paper.

\section{Reduction of the decay amplitude to Master Integrals}
\label{sec:MIReduction}
\setcounter{equation}{0}

\begin{figure}
\begin{center}
\includegraphics[width=8.3cm]{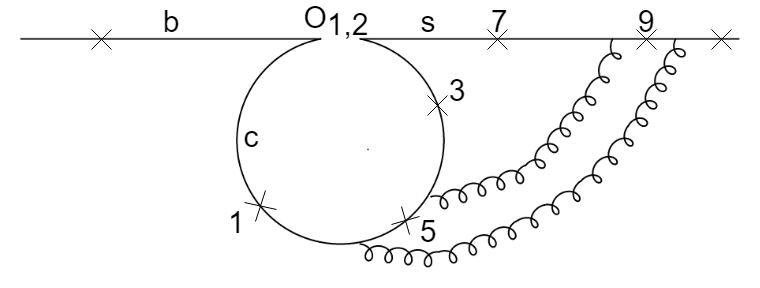}
\includegraphics[width=8.3cm]{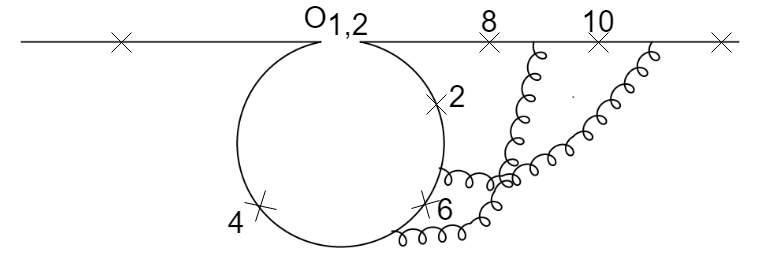}
\end{center}
\begin{center}
\includegraphics[width=8.3cm]{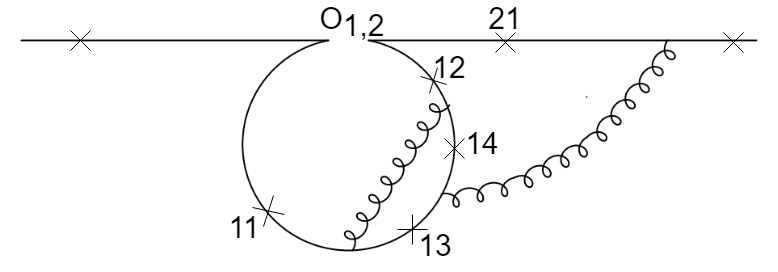}
\includegraphics[width=8.3cm]{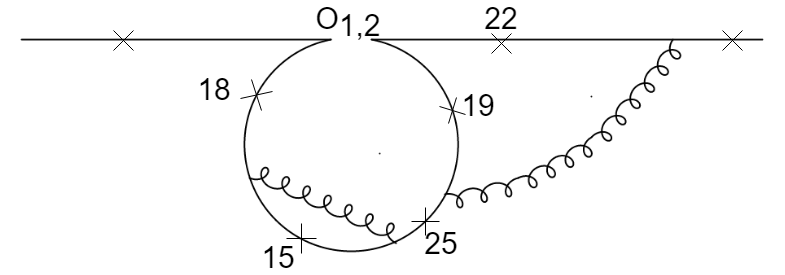}
\end{center}
\begin{center}
\includegraphics[width=8.3cm]{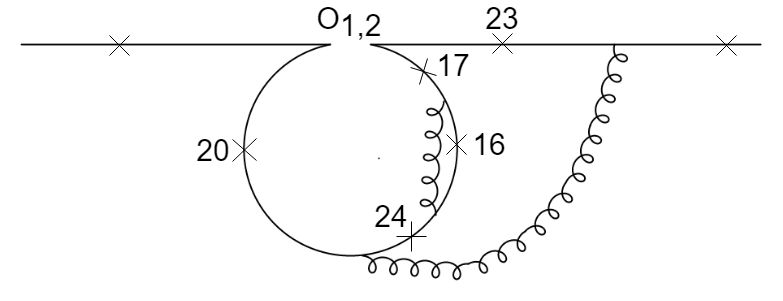}
\includegraphics[width=8.3cm]{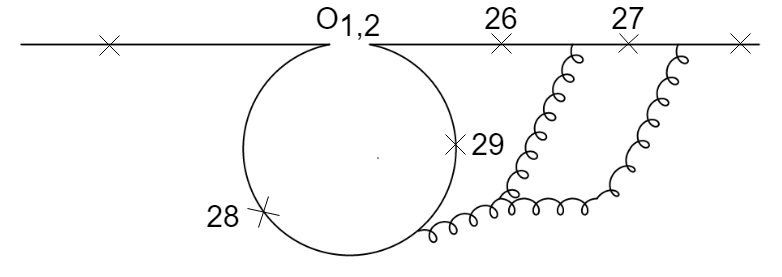}
\end{center}
\begin{center}
\includegraphics[width=8.3cm]{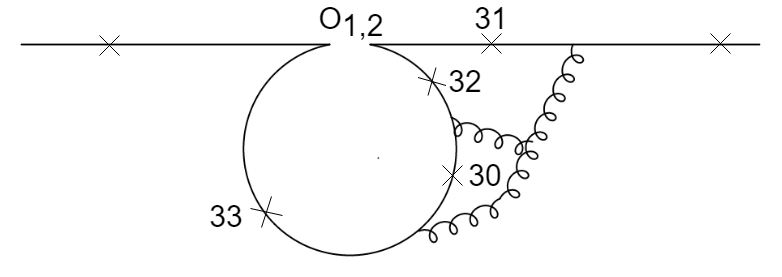}
\includegraphics[width=8.3cm]{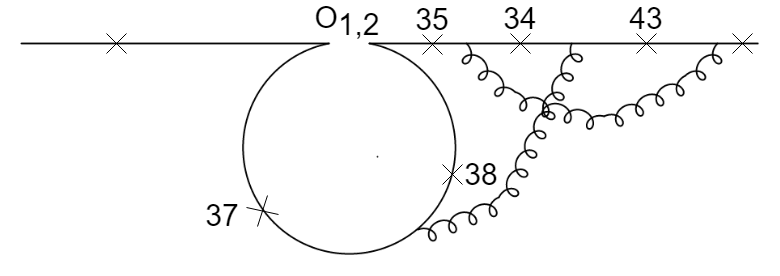}
\end{center}
\begin{center}
\includegraphics[width=8.3cm]{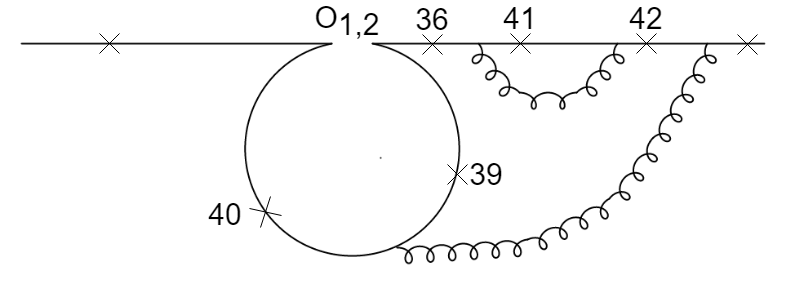}
\includegraphics[width=8.3cm]{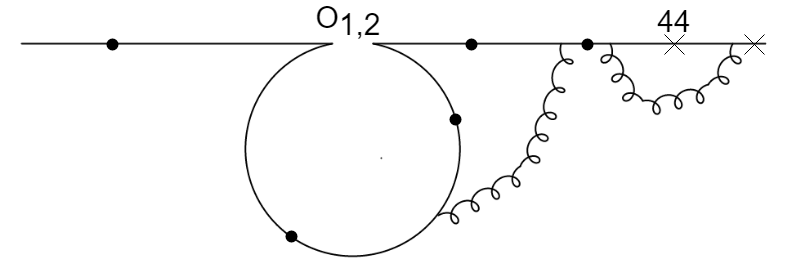}
\end{center}
\caption{List of those three-loop contributions to $b \to s \gamma$ associated with $O_1$ and $O_2$ which are worked out in this paper. A cross or a dot on a quark line represents a possible place where the photon can be emitted.
  While the diagrams marked by a dot will be taken into account in a future work (in connection with renormalization),
  we worked out the effects on the $b \to s \gamma$ decay amplitude of all diagrams with numbered and unnumbered crosses
  in this figure. To this end, as discussed in detail in section \ref{sec:3.1}, only the diagrams with numbered
  crosses had to be calculated explicitly.}
\label{fig:diagsleg}
\end{figure}

In this section we briefly describe the {\it algebraic steps} which reduce
the evaluation of a given three-loop diagram in \Fig{fig:diagsleg}
to the computation of so-called Master Integrals (MIs).
To this end, we decompose in section \ref{sec:3.1} the individual diagrams into form factors which can finally be represented as a linear combination of Scalar Integrals (SIs). In section \ref{sec:IBP}
we decompose these SIs into MIs. In section \ref{sec:DE} we
derive differential equations which govern the dependence on the charm quark mass of the MIs.
In section \ref{sec:MICalculation} we briefly discuss our intention of how to calculate the MIs appearing in the various diagrams.

\subsection{Form factor decomposition and scalar integrals}
\label{sec:3.1}

Using the standard Feynman rules in momentum space, the contribution to the decay amplitude ${\cal A}(b \to s \gamma)$ from a given Feynman diagram
$i$
in~\Fig{fig:diagsleg}
can be written as ${\cal A}^{(i)}=M_{\mu}^{(i)} \, \varepsilon^\mu$, where  $\varepsilon^\mu$ denotes the polarization vector of the emitted photon. After performing purely algebraic manipulations
like reductions of tensor integrals to scalar integrals (or alternatively using appropriate projection techniques) and freely using the equations of motion of the quark spinors and putting the strange quark mass to zero ($m_s=0$), $M_{\mu}^{(i)}$ can be written in the form 
\eq{
M_{\mu}^{(i)} =  \bar u_s (p_s) P_R  \left[ A^{(i)}\, q_\mu + B^{(i)}\, p_\mu + C^{(i)}\, \gamma_\mu \right] u_b(p)\ . \label{Mmui}
}
In this equation $u_b(p)$ denotes the Dirac spinor of the $b$-quark with four-momentum $p$, $u_s(p_s)$ is the analogous quantity for
the $s$-quark with four-momentum $p_s$ and $q=p-p_s$ is the four-momentum of the emitted photon. At this level, the form factors  $A^{(i)}$, $B^{(i)}$, $C^{(i)}$
are given in terms of linear combinations of scalar three-loop integrals. After performing these integrals and taking into account that $q^2=0$ in our process,
these form factors are functions of $m_b$ and $m_c$ (and depend also on the renormalization scale $\mu$).

Consider now a sum of diagrams which is gauge invariant with respect to QED. The quantity $M_{\mu}$ corresponding to this sum
then satisfies $q^\mu \, M_{\mu}=0 $, leading to the relation $C=-\frac{m_b}{2} \, B$, where $B$ and $C$ denote form factors of the this sum.
Using this relation, the corresponding gauge invariant amplitude  ${\cal A}=M_{\mu} \varepsilon^{\mu}$ can then be written as 
\eq{
  {\cal A} = -\frac{4 \pi^2}{e \, m_b} \, B \, \langle {O}_7 \rangle_{\rm tree} \, , \label{amplA}}
where the tree-level matrix element of the operator ${O}_7$ reads
\eq{
   \langle {O}_7 \rangle_{\rm tree} = \frac{e \, m_b}{16 \pi^2} \,  \bar u_s (p_s) P_R \left[ -4 p_{\mu} + 2 m_b \gamma_{\mu} \right] u_b(p) \, \varepsilon^{\mu} \, . }
Form eq. (\ref{amplA}) we see that a gauge invariant amplitude ${\cal A}$ can be written in such a way that only the form factor $B$ appears. As the aim of our paper is
to calculate the sum of all the diagrams in~\Fig{fig:diagsleg},
which is gauge invariant, it is sufficient to calculate only the form factors
$B^{(i)}$ of each individual diagram $i$. It is easy to see that those diagrams in~\Fig{fig:diagsleg} which are marked with a cross that does not
carry a number, only would contribute to the form factors $C^{(i)}$. Therefore only the diagrams with numbered crosses (1-44) have to be
worked out. A remark concerning the diagrams, where the photon emission is marked
by a ``dot'', is in order: We will take into account these contributions when working out the 2-loop counter-terms in a further publication. 

We now discuss how to evaluate the scalar quantities $B^{(i)}$ for the three-loop diagrams listed in~\Fig{fig:diagsleg}. 
To have a concrete example let us have a look at
diagrams 1 and 2 for which the momentum routings can be chosen such that
the same propagators (i.e. denominators) appear which allows to sum these two diagrams from the very beginning.
The results for the functions $B^{(i)}$ are given in terms of dimensionless three-loop scalar integrals of the type:
\eq{
j[n_{1},...,n_{12}] = (2\pi)^{-3d}
\int \frac{(m_b^2)^{N-6} (\tilde \mu^2)^{3\epsilon}\ d^d\ell \ d^dr_1 \ d^dr_2}{P_{1}^{n_{1}} P_{2}^{n_{2}} P_{3}^{n_{3}}P_{4}^{n_{4}} P_{5}^{n_{5}}P_{6}^{n_{6}}P_{7}^{n_{7}}P_{8}^{n_{8}}P_{9}^{n_{9}}P_{10}^{n_{10}}P_{11}^{n_{11}}P_{12}^{n_{12}}}
\label{eq:prop}}
where the numbers $n_i$ are integers (positive or negative), with $N= \sum_{i=1}^{12} n_{i}$, the objects $P_i$ are propagators (see below).
In addition, $d=4-2\epsilon$, and $\tilde \mu^2 \equiv \mu^2 e^{\gamma_E}/4\pi$,
with $\mu$ the $\overline {\rm MS}$ scale.
Our choice of momentum routings fixes the first eight propagators, while the other four (sometimes called artificial propagators) are chosen such that the twelve propagators form a linearly-independent set (in the sense that all occuring scalar products involving only loop-momenta or scalar products between loop-momenta and external momenta can be written as a linear combination of these propagators).
Explicitly, the complete list of propagators needed for diagrams 1 and 2 reads:
\begin{align}
P_1 &= (\ell+q)^2 - m_c^2 \, , & 
P_2 &= \ell^2 - m_c^2 \, ,& 
P_3 &= (\ell +r_2)^2 -m_c^2 \, , &
\nonumber\\
P_4 &=(\ell +r_1 +r_2)^2 - m_c^2 \, , & 
P_5 &= r_1^2 \, ,  &  
P_6 &= r_2^2 \, ,  &
\label{eq:propexpl1and2} \\ 
P_7 &= (r_2+p_s)^2  \, , &
P_8 &= (r_1 + r_2+p_s)^2 \, , &
P_9&=  (r_1+r_2)^2 \, , &
\nonumber \\
P_{10} &= (\ell+p_s)^2 \, , & 
P_{11} &= (r_1+q)^2  \, , &  
P_{12} &= (r_2+q)^2 \, . &
\nonumber
\end{align}

Returning to the complete list of diagrams in~\Fig{fig:diagsleg}, we define the following sets of diagrams,
in order to economize our calculation:
$(1,2)$, $(3,4)$, $(5,6)$, $(7,8)$, $(9,10)$, $(11,12)$, $(13,14)$, $(15, 16, 17, 18)$, $(19,20)$, $(21, 22, 23)$, $(24, 25)$, $(26)$, $(27)$, $(28, 29)$, $(30)$, $(31)$, $(32, 33)$, $(34)$, $(35, 36)$, $(37, 38, 39, 40)$, $(41, 42)$ and $(43, 44)$.
In a given set the momentum routings can be chosen in such a way that the same propagators appear, allowing to sum the diagrams
in this set from the very beginning. In the ancillary file submitted together with the present paper, we will give for each set $j$ the
corresponding form factor contribution $B^{(j)}$.
In the following we describe the analytic calculation of the three-loop scalar integrals. 

\subsection{IBP reduction and Master integrals}
\label{sec:IBP}

At this point, the result for a given set $j$ of diagrams is a linear combination of many scalar integrals. We can now apply integration-by-parts
identities (IBPs)
to reduce the scalar integrals to a small set of \emph{Master Integrals} (MIs).
For this purpose we use the code KIRA~\cite{KIRA,Usovitsch:2020jrk}
which implements Laporta's
algorithm~\cite{Laporta:2000dsw}. In earlier stages of our work also the program
LiteRed \cite{LightRed} was used. 

After reduction, we are left with $N$ three-loop MIs for the considered
set. For example, for set $(1,2)$ we get the MIs $J_i$ ($i=1,...,27$), which are listed in~\App{app:MIs}. 

We now perform some simplifying operations on the master integrals.
First, we express the integrands in terms of the dimensionless variable
\eq{
z \equiv m_c^2/m_b^2\ .
}

Second, in order to do a rational transformation to a canonical basis of master integrals whenever possible, we make a change of variable
$z \mapsto x$,
where $x = x(z)$ is a function to be specified later.
In terms of this new variable the dimensionless MIs are denoted as
$J_{i}(\epsilon,x)$.

\subsection{Differential equations for the Master integrals}
\label{sec:DE}
For a given set we first construct the system of differential equations for the MIs:
\eq{
\partial_x\,J_{k}(\epsilon,x) =
a^{k\ell}(\epsilon,x)\, J_{\ell}(\epsilon,x)\ ,
}
where ${\bf a}$ is a $N \times N$ matrix
depending on $\epsilon$ and $x$.
The derivatives of the MIs $J_{k}$ are performed by differentiating the integrands, which produce new scalar integrals, and then applying the IBP reduction again on these scalar integrals to express the derivatives $\partial_{x}\,J_{k}$ themselves in terms of the MIs $J_{k}$. One can then read off the matrix ${\bf a}$.

\subsection{Methods used to calculate the Master Integrals}
\label{sec:MICalculation}

For all sets except $(11,12)$ and $(13,14)$, we managed to transform the corresponding differential equations into canonical form. These canonical equations can then
be iteratively solved as an expansion in the dimensional regulator $\epsilon$.
In these solutions the dependence on the charm-quark mass is contained in
Generalized Polylogarithms (GPLs). In each iteration step new integration constants come into the game. The details for computation of the MIs for these diagrams (including comments about fixing the integration constants) are presented in section \ref{sec:PartI}. We stress that this method is purely analytic and therefore the preferred one.

For the MIs in sets $(11,12)$ and $(13,14)$, however, we did not manage to find a transformation to a canonical basis. We therefore used another method by performing a series expansion at $z=0$ (i.e., in terms of powers and logarithms of $z=m_c^2/m_b^2$). This method is described in some details in section \ref{sec:PartII}.

\section{Solving MIs via canonical form}
In this section we explain the details of the calculation of the MIs whose differential equations can be
transformed to canonical form (i.e., the MIs of all sets except  $(11,12)$ and $(13,14))$.
\label{sec:PartI}
\setcounter{equation}{0}

\subsection{Canonical form and iterative solution}
\label{sec:Canonical}
Adapted to our situation where the MIs depend (besides the dimensional regulator $\epsilon$) only on one variable $x$,
a basis of MIs is said to be ``canonical''~\cite{Henn:2013pwa} if ${\bf a}(\epsilon,x) = \epsilon {\bf A}(x)$, with ${\bf A}(x)$
being a $N \times N$ matrix independent of $\epsilon$. Given a canonical basis $\vec M$, the differential equation
has the form\footnote{A similar presentation is given in our paper \cite{Virto} for the case where the MIs depend on two variables $x$ and $y$.}:
\eq{
\partial_x \vec M(\epsilon,x) = \epsilon\  {\bf A}(x)\, \vec M(\epsilon,x) \quad . 
}

\bigskip

Once a canonical basis is found, the system of differential equations
can be solved automatically order by order in $\epsilon$.
To keep the notation as simple as possible in this section, we will assume that all the master integrals in the canonical basis are regular in $\epsilon$ (if not, we redefine them by multiplying all of them with the same appropriate power of $\epsilon$).
We then write the $\epsilon$-expansion for the master integrals
\eq{
\vec M(\epsilon,x) = \sum_{n=0}^\infty \epsilon^n\, \vec M_n(x) 
\label{eq:Mepsexp}}
and the differential equation reads:
\eq{
\partial_{x} \vec M_n(x) = {\bf A}(x) \vec M_{n-1}(x) \ .
\label{eq:diffeqcanonical}
}
Using partial fraction decomposition, ${\bf A}$ can be written in the form 
\eq{
{\bf A}(x) = \sum_j  \frac{{\bf A}^j}{x-w_j}\ ,
}
where ${\bf A}^j$ is a set of constant matrices, and the quantities $w_j$ are called the ``weights", which in our application
are just constants.

The differential equation (\ref{eq:diffeqcanonical}) can be solved iteratively: As $\vec M_{-1}(x)=0$ (see eq. (\ref{eq:Mepsexp}) and the text before this equation), (\ref{eq:diffeqcanonical}) implies that $\vec M_{0}(x)$ is constant, i.e., $\vec M_0(x)  =  \vec C_0$.
From $\vec M_0(x)$, we can get $\vec M_1(x)$ by just integrating (\ref{eq:diffeqcanonical}) for $n=1$ with respect to $x$. This step brings in a new integration constant $\vec C_1$. Repeating this procedure, we get 
\eqa{
\vec M_0(x) & = & \vec C_0 \ , 
\nonumber\\[2mm]
\vec M_1(x) & = &
\sum_{j_1} \big[{\bf A}^{j_1}\, G(w_{j_1};x)\big] \,\vec C_0 + \vec C_1 \ ,
\nonumber\\[2mm]
\vec M_2(x) & = &
\sum_{j_2,j_1} \big[{\bf A}^{j_2} \, {\bf A}^{j_1} \, G(w_{j_2},w_{j_1};x)\big] \, \vec C_0  +  \sum_{j_2} \big[{\bf A}^{j_2}\, G(w_{j_2};x)\big]  \, \vec C_1 + \vec C_2 \ ,
\nonumber\\[2mm]
\vec M_3(x) & = & \cdots
}
etc., in terms of Generalized Polylogarithms (GPLs)~\cite{Goncharov:1998kja}, defined iteratively as~\cite{Frellesvig:2016ske}
\eq{
G(w_1,\dots,w_n;x)=\int_0^x \frac{dt}{t-w_1} G(w_2,\dots,w_n;t)\ ;\quad
G(;x)=1\ ;\quad
G(\vec 0_n;x) = \frac{\log^n x}{n!}\ ,
}
where $\vec 0_n$ denotes $n$ consecutive zeroes.

Thus, the problem of calculating the MIs is reduced to find a canonical basis and to fix the integration constants, which is a much more tractable challenge.
In order to find canonical master integrals, we used the mathematica program {\it CANONICA}~\cite{Meyer:2017joq} (and for some checks also the program Libra \cite{Lee:2020zfb}).
The {\it CANONICA} code is able to look for transformations that involve \emph{rational} functions of the argument $x$. For this reason, the ``right'' variable $x$ must be found before using this program. Starting from our original variable $z=m_c^2/m_b^2$, we define $x$ as
\eq{x=\frac{1}{\sqrt{1-4z}}  \label{eq:xvar}}
in all sets considered in this paper.
In terms of this variable and with the help of {\it CANONICA}, we are able to find linear transformations
\eq{
M_{k} = (T^{-1})^{k\ell}(\epsilon,x)\, J_{\ell} \, ,
\label{eq:T}
}
such that the MIs $M_{k}$ constitute a canonical basis.
The weights occuring in the considered sets read:
\eq{
w_0=0, \, w_1=1, \, w_2=-1, \, w_3=\frac{i}{\sqrt{3}}, w_4=-\frac{i}{\sqrt{3}}, w_5=\frac{1}{\sqrt{5}}, w_6=-\frac{1}{\sqrt{5}} \, .}

We stress that the chosen variable $x$ has the property that it tends to zero when $z$ goes to infinity.
In this limit, the functions $G(...;x)$ can be expanded in a straightforward way for small values of $x$. 
This turns out to be very useful when fixing the integration constants in the following section, because we will heavily make use of the asymptotic properties of the
originals integrals $J_{k}$ in the limit where $x$ goes to zero.

\subsection{Fixing integration constants}
\label{sec:BoundaryConditions}

Once the canonical basis is found for a given set of diagrams and the general solution of the differential equations in this basis is constructed, we have to fix the integration constants. 
To this end we transform in a first step the MIs $\vec{M}$ back to the original basis consisting of the MIs $J_{k}$ by making use of the transformation matrix~$T$ (i.e.~\Eq{eq:T}).
The constants are then determined by computing some of the simpler MIs $J_{k}$ in a traditional way (i.e. through Feynman parametrization), while for the more difficult MIs
asymptotic properties in the limit $z \to \infty$ can be worked out. These properties follow in a straightforward way from the heavy mass expansion (HME) of a given integral \cite{Smirnov:1994tg}.

To be somewhat more concrete, we briefly explain which properties/statements were used to fix the constants for the 27 MIs appearing in the set $(1,2)$ (see equations (\ref{eq:appb}) and (\ref{eq:propexpl1and2})): 
\begin{enumerate}
\item
$J_1$ and $J_2$ can easily be worked out traditionally.
\item
  In the limit for large $m_c$ ($m_c \gg m_b$) the other 25 integrals can be naively Taylor expanded in the external momenta and in $m_b$. Note that in set $(1,2)$
  the only contributing subdiagrams of the MIs in the sense of the HME are just the full diagrams (i.e. the full MIs) and therefore the naive Taylor expansion is justified. The leading power (i.e. the maximal power) in the large $m_c$-expansion of a given integral $J_k$ is then of the form
$m_c^{n-6\epsilon}$, where the integer $n$ is identical to the
  mass dimension of the integral (strictly speaking of the integral in which the factor $(m_b^2)^{N-6}(\tilde \mu^2)^{3\epsilon}$ in the definition (\ref{eq:prop}) is understood to be omitted);
the structure of $J_k$ is
\eqa{
J_k=K \, m_c^{n-6\epsilon}\, P(m_b^2/m_c^2) \, , \label{eq:structurehme}}
where $K$ is a constant prefactor (w.r.t. $m_c$) and $P$ is a polynomial of the indicated argument.
The GPLs in the general solution for the MIs (from the differential equations) can easily be expanded for large $z$. Very often, the expanded solution for a given integral contains higher powers in $m_c$ than that determined from the HME argumentation. The requirement that these terms are absent allows to determine some of the integration constants. From the HME structure it is also clear that $n$ in equation (\ref{eq:structurehme}) is an {\it even} integer; this information again fixes some of the integration constants.
\item
The leading power in $m_c$ of the MI $J_6$ (which scales like $m_c^{4-6 \epsilon}$ in the large $m_c$ limit) coincides with $J_2$.
\item
The HME of the MI $J_{23}$ produces only the following powers of $m_c$: $m_c^{n-6\epsilon}$, where $n$ is an integer in the range $n={2,0,-2,-4,...}$. Therefore, when multiplying
$J_{23}$ with  $m_c^{+6 \epsilon}$ and then expanding in $\epsilon$, there can be no logarithms in $m_c$. This property fixes the remaining constants in set $(1,2)$. 
\end{enumerate}
It is worth emphasizing that in set $(1,2)$ all constants can be fixed by the explicit knowledge of $J_1$ and $J_2$ and structural information resulting from HME on the (integer parts) of the
powers of $m_c$ and on the logarithms in $m_c$.
{\it The explicit HME evaluation of the MIs is not even necessary.}

For many other sets of diagrams the fixing of the integration constants works in a similar way. However, in some sets the HME of certain MIs requires
to analyze genuine subdiagrams which make the extraction of the constants somewhat more complicated. Furthermore, in set $(21,22,23)$ a three-loop tadpole integral with four charm-lines appears which we could not calculate analytically; we took the results from \cite{Kniehl}
where this integral was calculated numerically to very high precision. 
To close this section, we note that our final results for all MIs have been checked numerically using Sector Decomposition as implemented in SecDec~\cite{SecDec} or PySecDec~\cite{pySecDec}.

\subsection{Expansion around $z=0$}
\label{sec:canonicalsmall}
The MIs of all sets (except $(11,12)$ and $(13,14)$) are now given in terms of GPLs and all the integrations constants
are fixed. However, the corresponding expressions are lengthy and the numerical evaluation (for example using GiNaC \cite{Vollinga:2005pk,ginac,Vollinga:2004sn}) of the large number of GPLs is time-consuming.
We therefore decided to expand all GPLs around $z=0$ (leading to powers and logarithms of $z$). Needless to say, the physical value of $z \sim 0.1$ is considerably
smaller than $0.25$, corresponding to the radius of convergence of this expansion. Using the expanded versions of the MIs, we worked out the form factor contributions $B^{(i)}$
for the individual sets $i$ of diagrams. The results for these $B^{(i)}$ quantities, consistently expanded up to the power $z^{10}$ (and including all power of $\log(z)$),
can be found in the mathematica file ``ancillary.m'' which is submitted together with this paper (for details see Appendix \ref{app:Bi}).

\section{Calculating the MIs in sets $(11,12)$ and $(13,14)$}
\label{sec:PartII}
\setcounter{equation}{0}
\subsection{Solving the differential equations as an expansion for small $z$}
\label{sec:smallz}
As we already mentioned in section \ref{sec:MICalculation},
we could not find transformations of the MIs $J_k$ to a canonical
basis for the sets $(11,12)$ and $(13,14)$, and therefore we propose another method for these two sets. As the physical value of $z=m_c^2/m_b^2$ is a small parameter (actually of order $0.1$), we construct
a series expansion for the solutions around $z=0$.
We start with the differential equation in matrix form as
\eq{\partial_z \vec J(\epsilon,z) = A(\epsilon,z) \vec J \, ,}
where $A(\epsilon,z)$ is an $N \times N$ matrix ($N$ is the number of MIs) which depends on $z$ and $\epsilon$ in a rational way.
First of all we transform the differential equation into Fuchsian form (see e.g. \cite{Lee:2020zfb}), i.e. in such a way that the transformed version of $A$
has at most $1/z$ singularities in all entries.
This can be achieved by transforming $\vec J \mapsto \vec J^{\, '}= L^{-1} \vec{J}$ with a diagonal matrix $L$ with entries of the form $L_{ii}=z^{n_i}$, where $n_i$ are suitably
chosen integers. The transformed matrix, denoted by $A_1$ then reads
\eq{A_1=\frac{\partial L^{-1}}{\partial z} L + L^{-1} A L \, .}
By construction, the singular part of $A_1$, denoted by $A_s$, is then proportional to $\frac{1}{z}$.
In the next step we apply a further transformation $\vec J^{\, '} \mapsto \vec J^{\, ''} = S^{-1} \vec J^{\, '}$ which brings $A_s$
to Jordan form, i.e. to a matrix with upper triangular form, for which only the elements on the diagonal and above the diagonal are nonzero.
Practically, this is done in Mathematica, using the command $S = JordanDecomposition[A_s][[1]]$. Note that $S$ depends on $\epsilon$, but not on $z$.
The MIs $\vec J^{\, ''}$ then obey the differential equation
\eq{ \partial_z \vec J^{\, ''} = A_2(\epsilon,z) \vec J^{\, ''} \quad \mbox{with} \quad A_2=S^{-1} A_1 S  \, . \label{eq:eqA2}}
This first-order linear differential equation, which we now want to solve, will have $N$ linearly independent fundamental solutions: $\vec J^{\, ''}_1,...,\vec J^{\, ''}_N$.  
We first construct the leading part of these solutions which correspond to take the $\frac{1}{z}$ part of $A_2$ in (\ref{eq:eqA2}).
We stress that these parts can be easily obtained (e.g. in Mathematica), because the leading part of $A_2$ is by construction in Jordan form. It is instructive
to explicitly display the leading part for a few of the $N$ solutions
as they appear when calculating the MIs
for set $(11,12)$ (where the number of MIs is 31). Using $\ell=\log(z)$, we have:
\eqa{
  \vec J^{\, ''}_{1,{\rm lead}} = &&z (1, 0, 0, 0, 0, 0, 0, 0, 0, 0, 0, 0, 0, 0, 0, 0, 0, 0, 0, 0, 0, 0, 0,
0, 0, 0, 0, 0, 0, 0, 0) \nonumber \\
\vec J^{\, ''}_{5,{\rm lead}} = && z^{2 - 3 \epsilon} (0, 0, 0, 0,  \ell,1, 0, 0, 0, 0, 0, 0, 0, 0, 0, 0, 0, 0, 0, 0, 0, 0, 0, 0, 0, 0,
0, 0, 0, 0, 0) \nonumber \\
\vec J^{\, ''}_{6,{\rm lead}} = && z^{2 - 3 \epsilon} (0, 0, 0, 0, 1, 0, 0, 0, 0, 0, 0, 0, 0, 0, 0, 0, 0, 0, 0, 0, 0, 0, 0, 0, 0, 0, 0, 0, 0, 0, 0) \nonumber \\
\vec J^{\, ''}_{9,{\rm lead}} = && z^{3 - 2 \epsilon}(0, 0, 0, 0, 0, 0, 0, 0, 1, 0, 0, 0, 0, 0, 0, 0, 0, 0, 0, 0, 0, 0, 0, 0, 0, 0, 0, 0, 0,
0, 0) \nonumber \\
\vec J^{\, ''}_{26,{\rm lead}} = && z^{3 - 3 \epsilon} (0, 0, 0, 0, 0, 0, 0, 0, 0, 0, 0, 0, 0, 0, 0, 0, 0, 0, 0, 0, 0, 0, 0,
\frac{\ell^2}{2}, \ell, 1, 0, 0, 0, 0, 0) \nonumber \\}
We see that the leading solutions are proportional to $z^{(n_0-m_0\,\epsilon)}$ and also involve powers of $\log(z)$. When taking into account all the $N$ fundamental solutions, we find that 
$n_0=1,2,3$, $m_0=0,1,2,3,4$ and the maximal power of $\log(z)$ is 2. In order to get the subleading terms
of the fundamental solutions (i.e. higher powers in $z$), we make an ansatz: When
the leading power of a given fundamental solution is proportional to $z^{n_0-m_0\,\epsilon}$, we
add to $i$-th component of the leading part the following term
\eq{\sum_{n=n_0+1}^{n_{\rm max}}{z^{n-m_0\,\epsilon}(c^0_{i,n} +  c^1_{i,n} \log(z) +    c^2_{i,n} \log^2(z))} \, ,}
where the quantities $c^{0,1,2}_{i,n}$ denote $\epsilon$-dependent coefficients. 
We then insert this ansatz into the differential equation (\ref{eq:eqA2}) and expand the left- and right-hand side up to order $n_{\rm max}-1$.
Requiring the corresponding powers of $z$ and $\log(z)$ to be equal, leads to a system of linear
algebraic equations for these coefficients, which can be solved directly. In such a way we get $N$ linear independent fundamental solutions $\vec J^{\, ''}_1,...,\vec J^{\, ''}_N$ which contain powers up to $z^{n_{\rm max}-m_0\,\epsilon}$ together with
$\log(z)$-terms up to second power.
Then we transform back each of these $N$ fundamental solutions to the original basis by doing the inverse transformations with $S$ and $L$, leading
to the $N$ fundamental solutions $\vec J_1,...,\vec J_N$. The original MIs $\vec{J}$, which we are finally interested in, are then linear combinations of the $N$ fundamental solutions, i.e.,
\eq{\vec{J} = \sum_{i=1}^N{C_i(\epsilon) \vec J_i} \, . \label{eq:lincombJi}}
The $N$ coefficients  $C_i(\epsilon)$ play the role of integration constants, which we will fix below by exploiting suitable information extracted from
the integral representations of the MIs near $z=0$.
We note that so far no expansion in $\epsilon$ was performed. 
Also note that the MIs $\vec{J}$ are known with a maximum power of $z$ which is less than $n_{\rm max}$, because the back transformation matrix $L$
contains terms proportional $\frac{1}{z^i}$ with positive $i$.
\subsection{Fixing the integration constants $C_i(\epsilon)$}
\label{sec:initsmallz}
We briefly describe how to fix the integration constants $C_i(\epsilon)$ which
appear in \Eq{eq:lincombJi}. To this end, it is convenient to cast this equation in component form:
\eq{J_k = \sum_{i=1}^N{C_i(\epsilon) J_{i,k}} \, . \label{eq:lincombJik}}
Note that the $\epsilon$ dependence of the fundamental solutions $J_{i,k}$ on the right hand side of this equation is still exact (i.e. not expanded), while 
the $z$-dependence is contained in terms of the form
\eq{z^{n-m_0 \epsilon} \, , \quad  z^{n-m_0 \epsilon} \, \log(z) \, , \quad  z^{n-m_0 \epsilon} \, \log^2(z)\, , }
where $n$ is a non-negative integer and $m_0=0,1,2,3,4$. 
For each $J_k$ we now build a leading power version $J_k^{{\rm lead}}$
by keeping for any given $m_0$ only all the terms with the smallest $n$.

To gain information on the integration constants, we use {\it the method of regions} (see e.g. \cite{Beneke:1997zp,Smirnov:book1,Smirnov:book2}),
in particular the version formulated in Feynman- or $\alpha$-parameter space as described in sections 9.2 and 9.3 in \cite{Smirnov:book1,Smirnov:book2}, respectively. This version is
implemented in the program FIESTA5 \cite{Smirnov:2021rhf}
(which we mainly called with the QMC integrator \cite{SBorowka}; for checks we also called FIESTA5 with the VEGAS integrator \cite{THahn}).
This program (after some minor adaptions done by us) allows to numerically calculate the leading versions of the MIs $J_k$ directly from their integral representations. For many MIs we could check specific regions even analytically by using the program HyperInt \cite{Panzer:2014caa}.

The requirement that the leading versions
of the left- and the right-hand side of \Eq{eq:lincombJik} coincide, fixes the integration constants. 
\subsection{Numerical consistency studies for set (11,12)}
\label{sec:numerics}
\begin{figure}
\begin{center}
\includegraphics[width=15.0cm]{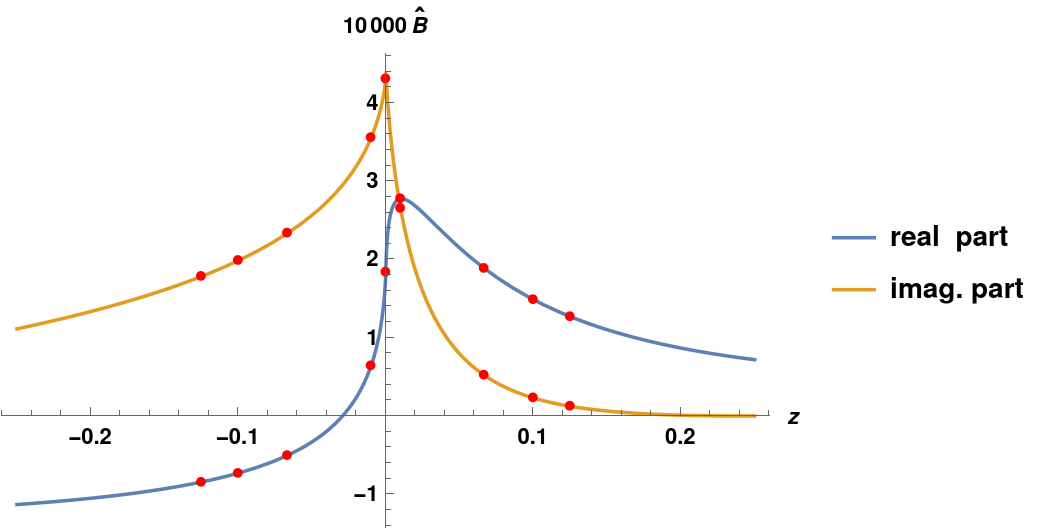}
\end{center}
\caption{Form factor contribution $10^4 \times \hat{B}$ as defined in \Eq{eq:Bhat}. See text for details.}
\label{fig:plot}
\end{figure}
In this section we investigate a few aspects of the form factor contribution $B_{(11,12)}$ corresponding to the set $(11,12)$.
We write this contribution as
\eq{B_{(11,12)} = {\rm pref} \cdot \hat{B} \quad \mbox{with} \quad {\rm pref}=-\frac{e \, m_b}{4 \pi^2} \, g_s^4\,\left( \frac{\mu^2}{m_b^2} \right)^{3 \epsilon} \, . \label{eq:Bhat}}
\Fig{fig:plot} shows the dimensionless quantity $\hat{B}$, where the solid curves are obtained as described in
sections \ref{sec:smallz} and \ref{sec:initsmallz}. The results for negative values of $z$
are related to those for positive $z$ (i.e. physical values) via analytic continuation
through the lower $z$-half-plane. As a consistency check, we calculated $\hat{B}$
for the values $z=1/100$, $z=1/15$, $z=1/10$ and $z=1/8$ by directly working
out the 31 MIs numerically at these values for $z$ using SecDec and/or PySecDec (without using method by region features). These results are shown by red dots in \Fig{fig:plot}. Furthermore,
we did a similar direct numerical calcuation using FIESTA5 for the corresponding negative values for $z$ (again without using method by region features). These results are also shown by red dots in
\Fig{fig:plot}. Finally, we did a completely new calculation by putting $z=0$ from the very beginning.
In this case, the form factor contribution $\hat{B}$ is given in terms of only $3$ MIs, viz. $J_{7}$, $J_{18}$ and $J_{25}$ (see \Eq{eq:misappc}),
which we calculated for $m_c=0$ using
SecDec. 
The corresponding result is also show by a dot. We think that \Fig{fig:plot} nicely shows the intrinsic consistency of our approach.

\section{Result as a Power Series on $z$ and $\log(z)$}
\label{sec:Result}
In this paper we worked out the contributions to the $b \to s \gamma$ decay amplitude of all three-loop diagrams 
in \Fig{fig:diagsleg} which are marked with crosses (both numbered and unnumbered crosses).
We denote the
contributions related to the operators $O_1$ and $O_2$ as ${\cal A}_1$ and   ${\cal A}_2$, respectively. 
These diagrams form a gauge invariant set (w.r.t. to QED) and therefore,
as detailed in section \ref{sec:3.1},
it is sufficient to explicitly calculate the form factor $B$ (see \Eq{amplA}) to which only the diagrams
with numbered crosses contribute. While the form factor contributions $B^{(i)}$ from the individual sets of diagrams are given in electronic form in the file ``ancillary.m'' (see Appendix \ref{app:Bi}), we give
here the results for ${\cal A}_1$ and ${\cal A}_2$
as power series on $z$ where we list terms up to $z^5$. For a given power of $z$ we retain all powers of  $\log(z)$, which
practically means up to $\log^5(z)$ as there are no higher $\log$-powers. The results read (for $k=1,2$)
\eq{
  {\cal A}_k = g_s^4 \, \left(\frac{\mu^2}{m_b^2}\right)^{3\epsilon} \, \langle {O}_7 \rangle_{\rm tree}\,\left [ \frac{c^{(k,-2)}}{\epsilon^2}+\sum\limits_{i=0}^5\sum\limits_{j=0}^5 \left( \frac{1}{\epsilon} c^{(k,-1)}_{ij} + c^{(k,0)}_{ij} \right) z^i \log(z)^j  \right ]   
  \, , \label{eqn:zlogz}}
where  $c^{(1,-2)}=0.00002701 + 0.000006681\, z$,  $c^{(2,-2)}=-0.0001115 - 0.00004008\, z$, and $c^{(1,-1)}_{ij}$, $c^{(1,0)}_{ij}$, $c^{(2,-1)}_{ij}$, $c^{(2,0)}_{ij}$  are given in Tables \ref{tab:epso1m1},\ref{tab:epso10},\ref{tab:epso2m1} and \ref{tab:epso20}. We mention that the reader can easily obtain
results for ${\cal A}_1$ and ${\cal A}_2$
up to the $z^{10}$ by using the information in the file ``ancillary.m''. The results when going up to $z^{5}$ and those when going up to $z^{10}$ are close to each other, for instance for $z=1/8$ the relative difference is only of order $0.5\%$.

\begin{table}
\tiny
\hspace{-2.5cm}
\begin{center}
\begin{tabular}{|c|c|c|c|c|c|c|} 
\hline
 $O_1, \, \frac{1}{\epsilon}$  & $z^0$ & $z^1$ & $z^2$ & $z^3$& $z^4$& $z^5$ \\ [1ex] 
 \hline\hline
 $\log(z)^0$  & $0.298 + 0.255 I$& $ 1.047 - 1.616 I $&$
0.733 + 3.407 I $&$
-1.441 - 3.546 I$&$
-0.390 - 0.258 I$&$
-0.224 - 1.412 I$  \\  \hline

$\log(z)^1$  &  $0$&$

0.834 - 1.138 I$&$

2.870 - 0.550 I$&$

-1.268 + 1.300 I$&$

-0.0821 - 0.0111 I$&$

-0.449 - 0.0691 I$  \\  \hline

$\log(z)^2$  &  $0$&$

-0.240 - 0.653 I$&$

-0.0875 - 0.852 I$&$

0.207 + 0.0664 I$&$

-0.00176$&$

-0.0110$   \\  \hline

$\log(z)^3$  & $0$&$

-0.0693 + 0.0187 I$&$

-0.0904$&$

0.00704$&$

0$&$

0$  \\  \hline

$\log(z)^4$  & $0$ & $0.00148$  & $0$ & $0$ & $0$ & $0$   \\  \hline

$\log(z)^5$  & $0$ & 0  & $0$ & $0$ & $0$ & $0$    \\    \hline

\end{tabular}\\
\end{center}
\caption{Numerical values for $10^3 \times c^{(1,-1)}$ (see equation (\ref{eqn:zlogz})).}
\label{tab:epso1m1}
\end{table}

\begin{table}
\tiny
\hspace{-2.5cm}
\begin{center}
\begin{tabular}{|c|c|c|c|c|c|c|} 
 \hline
 $O_1, \, \epsilon^0$  & $z^0$ & $z^1$ & $z^2$ & $z^3$ & $z^4$ & $z^5$ \\ [1ex] 
 \hline\hline
$\log(z)^0$  & $0.0702 + 0.281 I$&$

1.248 - 0.655 I$&$

-3.087 + 1.928 I$&$

4.573 - 1.446 I$&$

-1.544 - 1.130 I$&$

2.270 - 1.244 I$ \\  \hline

$\log(z)^1$  &  $0$ &$0.444 - 0.0479 I$&$

1.128 + 0.958 I$&$

-1.392 - 1.805 I$&$

-0.0765 + 1.372 I$&$

-0.201 - 1.063 I$    \\  \hline

$\log(z)^2$  &  $0$ & $-0.033 - 0.158 I$&$

-0.257 - 0.162 I$&$

-0.0339 + 0.261 I$&$

0.360 - 0.0760 I$&$

-0.185 - 0.0401 I$  \\  \hline

$\log(z)^3$  & $0$  & $-0.001 + 0.099 I$&$

0.00390 + 0.164 I$&$

0.00704 - 0.0334 I$&$

-0.00789 - 0.0412 I$&$

-0.00315 + 0.0265 I$  \\  \hline

$\log(z)^4$  & $0$   &$0.0131 - 0.000829 I$&$

0.0198 - 0.00296 I$&$

-0.00318$&$

-0.00328$&$

0.00211$   \\  \hline

$\log(z)^5$  & $0$&$

-0.000142$&$

-0.000188$&$

0$&$

0$&$

0$  \\    \hline
\end{tabular}\\
\end{center}
\caption{Numerical values for $10^2 \times c^{(1,0)}$ (see equation (\ref{eqn:zlogz})).}
\label{tab:epso10}
\end{table}
\begin{table}
\tiny
\hspace{-2.5cm}
\begin{center}
\begin{tabular}{|c|c|c|c|c|c|c|} 
\hline
 $O_2, \,\frac{1}{\epsilon}$  & $z^0$ & $z^1$ & $z^2$ & $z^3$& $z^4$& $z^5$ \\ [1ex] 
 \hline\hline
 $\log(z)^0$  & $-1.227 - 1.051 I$&$

-4.920 + 7.970 I$&$

-4.460 - 17.128 I$&$

7.240 + 18.141 I$&$

2.066 + 1.436 I$&$

1.157 + 7.770 I$  \\  \hline

$\log(z)^1$  &  $0$ & $-3.441 + 5.822 I$&$

-14.052 + 3.298 I$&$

6.608 - 6.454 I$&$

0.457 + 0.0664 I$&$

2.473 + 0.415 I$  \\  \hline

$\log(z)^2$  &  $0$ &  $1.278 + 2.912 I$&$

0.525 + 4.106 I$&$

-1.027 - 0.398 I$&$

0.0106$&$

0.066$   \\  \hline

$\log(z)^3$  & $0$ & $0.309 - 0.112 I$&$

0.436$&$

-0.0422$ & $0$ & $0$  \\  \hline

$\log(z)^4$  & $0$ & $-0.0089$  & $0$ & $0$ & $0$ & $0$   \\  \hline

$\log(z)^5$  & $0$ & 0  & $0$ & $0$ & $0$ & $0$    \\    \hline

\end{tabular}\\
\end{center}
\caption{Numerical values for $10^3 \times c^{(2,-1)}$ (see equation (\ref{eqn:zlogz})).}
\label{tab:epso2m1}
\end{table}

\begin{table}
\tiny
\hspace{-2.5cm}
\begin{center}
\begin{tabular}{|c|c|c|c|c|c|c|} 
 \hline
 $O_2, \,\epsilon^0$  & $z^0$ & $z^1$ & $z^2$ & $z^3$ & $z^4$ & $z^5$ \\ [1ex] 
 \hline\hline
$\log(z)^0$  & $-0.330 - 1.157 I$&$

-2.922 + 2.250 I$&$

1.091 - 6.952 I$&$

0.0581 + 1.884 I$&$

-16.856 + 5.502 I$&$

3.643 + 10.932 I$ \\  \hline

$\log(z)^1$  &  $0$ &$-1.284 - 1.243 I$&$

-5.118 + 2.733 I$&$

-1.066 - 6.449 I$&$

1.349 + 11.105 I$&$

3.797 - 7.035 I$    \\  \hline

$\log(z)^2$  &  $0$ & $0.107 + 0.0505 I$&$

2.071 + 1.250 I$&$

-2.180 + 1.694 I$&$

2.752 - 0.124 I$&$

-2.469 - 0.441 I$  \\  \hline

$\log(z)^3$  & $0$  & $-0.120 - 0.515 I$&$

0.193 - 0.671 I$&$

0.282 + 0.131 I$&$

-0.0142 - 0.335 I$&$

-0.0534 + 0.311 I$  \\  \hline

$\log(z)^4$  & $0$   &$-0.0642 + 0.0162 I$&$

-0.0861 - 0.0271 I$&$

0.0136$&$

-0.0267$&$

0.0248$   \\  \hline

$\log(z)^5$  & $0$    &$0.00156$&$

-0.00172$     &$0$   &$0$   &$0$  \\    \hline
\end{tabular}\\
\end{center}
\caption{Numerical values for $10^2 \times c^{(2,0)}$ (see equation (\ref{eqn:zlogz})).}
\label{tab:epso20}
\end{table}

\section{Summary and outlook}
\label{sec:conclusions}
\setcounter{equation}{0}
In this paper we worked out three-loop diagrams (of order $\alpha_s^2$) contributing
to the decay amplitude for $b \to s \gamma$ associated with the current-current operators
$O_1$ and $O_2$ at the physical value of $m_c$. As the corresponding calculations are among the hardest we have ever done,
we concentrated only on the well-defined class of diagrams
where no gluon is touching the $b$-quark line (see \Fig{fig:diagsleg}) in this paper; note that we did not work out the diagrams with closed
fermion bubbles inserted into gluon-lines, because these contributions already exist in the literature.
For many diagrams we could solve the master integrals using differential equations in the canonical form, as presented in section
\ref{sec:PartI}. However, for the four diagrams $(11-14)$ we could not find a transformation to canonical form and we therefore calculated in section~\ref{sec:PartII}
the corresponding master integral directly as an expansion around $z=0$, retaining power terms up to $z^{10}$ and keeping the accompanying  $\log(z)$
terms to all powers. The results for the sum of all considered diagrams are given in tabular form in section~\ref{sec:Result} and also in electronic form in the file ancillary.m,
which is submitted together with this paper (see Appendix~\ref{app:Bi}).
Making use of the two methods discussed in this paper,
we are confident that we will manage to work out the remaing three-loop diagrams (at the physical value of $m_c$),
completing the virtual QCD corrections of order $\alpha_s^2$ to the decay amplitude ${\cal A}(b \to s \gamma)$ associated with
$O_1$ and $O_2$.

\section*{Acknowledgements}
\noindent
C.G. is very grateful to G. Heinrich, V. Magerya, J. Schlenk, and especially to S. Jones, for useful discussions on SecDec and PySecDec, in particular on the
``method by regions'' features. He also  acknowledges useful discussions with Christoph Meyer on his program CANONICA as well as discussions with Johann Usovitsch on
finding a ``good bases'' of master integrals with his program factorizeBasis.nb which extends KIRA. Further thanks go to R.N. Lee for questions on his programs LiteRed and Libra.
C.G. also would like to thank J. Gasser for lecturing (a long time ago) about Fuchsian equations in his courses on quantum mechanics. Also very
useful discussions with N. Schalch on various features connected with multiloop diagrams are greatfully acknowledged.

\noindent
The work of C.G., C.W. and F.S. is partially supported by the Swiss National Science Foundation under grants 200020-175449
and 200020-204075.

\noindent
H.M.A. is  supported by the Committee of Science of Armenia Program Grant No.
21AG‐1C084.

\newpage

\appendix
\renewcommand{\theequation}{\Alph{section}.\arabic{equation}} 

\setcounter{equation}{0}

\section{Details on the ancillary file}
\subsection{Results for the individual contributions to the form factor $B$ in electronic form}
\label{app:Bi}

In the mathematica-file ``ancillary.m'' (which is included in the submission of this paper) we give the contributions to the form factor $B$ (see \Eq{Mmui}) for the following 
$22$ sets of diagrams:
$(1,2)$, $(3,4)$, $(5,6)$, $(7,8)$, $(9,10)$, $(11,12)$, $(13,14)$, $(15, 16, 17, 18)$, $(19,20)$, $(21, 22, 23)$, $(24, 25)$, $(26)$, $(27)$, $(28, 29)$, $(30)$, $(31)$, $(32, 33)$, $(34)$, $(35, 36)$, $(37, 38, 39, 40)$, $(41, 42)$ and $(43, 44)$.
In this file, the contributions from sets $(1,2)$ and $(37,38,39,40)$ are denoted as ``B1to2'' and ``B37to40'', respectively (and so on). The expressions contain the results as an expansion in $z$ (around $z=0$), where terms up to $z^{10}$ are retained. Note that
for a given power of $z$ all powers of $\log(z)$ are kept. The formulas contain the symbolic prefactor ``pref'', which amounts to
\eq{{\rm pref}=-\frac{e \, m_b}{4 \pi^2} \, g_s^4\,\left( \frac{\mu^2}{m_b^2} \right)^{3 \epsilon} \, , }
as well as symbolic color-factors ``col[i]''. The actual values for these color-factor are also given in this file for both, the $O_1$- and the $O_2$-contributions; they are given as mathematica-lists denoted by ``coloro1'' and ``coloro2'',
being written in terms of the number of colors $N_c$ ($N_c=3$). If the form factor contributions related to $O_1$ are of interest,
the following replacement should be done (e.g. in the mathematica notebook into which the ancillary file is imported):
\eq{ {\rm col}[i_{-}] :> {\rm coloro1}[[i]] }
Furthermore the formulas contain the symbolic charge-factors $Q_c$ and $Q_s$ (whose numerical values are $Q_c=2/3$ and $Q_s=-1/3$).

According to section \ref{sec:3.1}, the decay amplitude corresponding to the diagrams considered in this paper (see \Fig{fig:diagsleg})
is obtained through
\eq{{\cal A}_k = - \frac{4 \pi^2}{e \, m_b} \, B_k \, \langle O_7 \rangle_{{\rm tree}} \qquad (k=1,2) \, , }
where $B_k$ is the sum of the $22$ form factor contributions associated with the operator $O_k$ ($k=1,2$).
In order to check that the ancillary file works properly, the reader is invited to reproduce the coefficents in \Eq{eqn:zlogz} (which are given in
Tables \ref{tab:epso1m1}, \ref{tab:epso10}, \ref{tab:epso2m1} and \ref{tab:epso20}).

\section{List of Master Integrals for diagrams 1 and 2}
\label{app:MIs}

In this appendix we list the Master Integrals (MIs) $J_{1},...,J_{27}$
which appear in the calculation of the three-loop diagrams 1 and 2 contained
in~\Fig{fig:diagsleg}.
The notation is described in equation (\ref{eq:prop})  and the explicit form of the propagators
is given in equation (\ref{eq:propexpl1and2}).

\bigskip

\noindent The 27 MIs read:
\begin{align}
& J_{1}=j[1, 0, 1, 1, 0, 0, 0, 0, 0, 0, 0, 0]
&& J_{2}=j[1, 0, 0, 1, 1, 1, 0, 0, 0, 0, 0, 0]
\nonumber\\
& J_{3}=j[1, 0, 1, 1, 0, 0, 1, 0, 0, 0, 0, 0]
&& J_{4}=j[1, 0, 1, 1, 0, 0, 2, 0, 0, 0, 0, 0]
\nonumber\\
& J_{5}=j[1, 1, 1, 1, 0, 0, 1, 0, 0, 0, 0, 0]
&& J_{6}=j[1, 0, 0, 1, 1, 0, 1, 0, 0, 0, 0, 0]
\nonumber\\
& J_{7}=j[1, 0, 0, 1, 1, 0, 2, 0, 0, 0, 0, 0]
&& J_{8}=j[1, 0, 0, 2, 1, 0, 1, 0, 0, 0, 0, 0]
\nonumber\\
& J_{9}=j[1, 1, 0, 1, 1, 1, 1, 0, 0, 0, 0, 0]
&& J_{10}=j[1, 0, 1, 1, 0, 1, 0, 1, 0, 0, 0, 0]
\nonumber\\
& J_{11}=j[1, 0, 1, 1, 0, 1, 0, 2, 0, 0, 0, 0]
&& J_{12}=j[1, 0, 1, 1, 0, 2, 0, 1, 0, 0, 0, 0]
\nonumber\\
& J_{13}=j[1, 0, 1, 2, 0, 1, 0, 1, 0, 0, 0, 0]
&& J_{14}=j[1, 0, 2, 1, 0, 1, 0, 1, 0, 0, 0, 0]
\label{eq:appb} \\
& J_{15}=j[1, 0, 0, 1, 1, 1, 0, 1, 0, 0, 0, 0]
&& J_{16}=j[1, 0, 0, 1, 1, 1, 0, 2, 0, 0, 0, 0]
\nonumber\\
& J_{17}=j[1, 1, 0, 1, 1, 1, 0, 1, 0, 0, 0, 0]
&& J_{18}=j[1, 0, 1, 1, 1, 1, 0, 1, 0, 0, 0, 0]
\nonumber\\
& J_{19}=j[1, 0, 1, 1, 1, 1, 0, 2, 0, 0, 0, 0]
&& J_{20}=j[1, 0, 1, 1, 1, 2, 0, 1, 0, 0, 0, 0]
\nonumber\\
& J_{21}=j[1, 0, 1, 1, 2, 1, 0, 1, 0, 0, 0, 0]
&& J_{22}=j[1, 1, 1, 1, 1, 1, 0, 1, 0, 0, 0, 0]
\nonumber\\
& J_{23}=j[1, 0, 1, 1, 0, 0, 1, 1, 0, 0, 0, 0]
&& J_{24}=j[1, 0, 1, 1, 0, 0, 1, 2, 0, 0, 0, 0]
\nonumber\\
& J_{25}=j[1, 0, 1, 2, 0, 0, 1, 1, 0, 0, 0, 0]
&& J_{26}=j[1, 1, 1, 1, 0, 0, 1, 1, 0, 0, 0, 0]
\nonumber\\
& J_{27}=j[1, 0, 1, 1, 0, 1, 1, 1, 0, 0, 0, 0]
&&
\nonumber
\end{align}

\section{List of Master Integrals for diagrams 11 and 12}
\label{app:MI1112s}

In this appendix we list the Master Integrals (MIs) $J_{1},...,J_{31}$
which appear in the calculation of the three-loop diagrams 11 and 12 contained
in~\Fig{fig:diagsleg}.
The set of propagators reads (where again the first eight are physical and the remaining four are artificial):
\begin{align}
P_1 &= \ell^2 - m_c^2 \, , & 
P_2 &= (\ell+q)^2 - m_c^2 \, ,& 
P_3 &= (\ell +r_1)^2 -m_c^2 \, , &
\nonumber\\
P_4 &=(\ell +r_2)^2 - m_c^2 \, , & 
P_5 &= (\ell +r_1+r_2)^2 - m_c^2 \, ,  &  
P_6 &= r_1^2 \, ,  &
\label{eq:propexpl11and12} \\ 
P_7 &= r_2^2  \, , &
P_8 &= (r_1 +p_s)^2 \, , &
P_9&=  (\ell + p_s)^2 \, , &
\nonumber \\
P_{10} &= (r_1+q)^2 \, , & 
P_{11} &= (r_2+q)^2  \, , &  
P_{12} &= (r_2+p_s)^2 \, . &
\nonumber
\end{align}

\bigskip

\noindent The 31 MIs read:
\begin{align}
& J_{1}=j[1, 0, 1, 1, 0, 0, 0, 0, 0, 0, 0, 0]
&& J_{2}=j[1, 0, 1, 1, 1, 0, 0, 0, 0, 0, 0, 0]
\nonumber\\
& J_{3}=j[0, 0, 1, 1, 0, 1, 1, 0, 0, 0, 0, 0]
&& J_{4}=j[0, 1, 1, 1, 0, 0, 0, 1, 0, 0, 0, 0]
\nonumber\\
& J_{5}=j[0, 1, 1, 1, 0, 0, 0, 2, 0, 0, 0, 0]
&& J_{6}=j[1, 1, 1, 1, 0, 0, 0, 1, 0, 0, 0, 0]
\nonumber\\
& J_{7}=j[0, 1, 1, 1, 1, 0, 0, 1, 0, 0, 0, 0]
&& J_{8}=j[0, 1, 1, 1, 1, 0, 0, 2, 0, 0, 0, 0]
\nonumber\\
& J_{9}=j[0, 1, 1, 1, 2, 0, 0, 1, 0, 0, 0, 0]
&& J_{10}=j[0, 1, 2, 1, 1, 0, 0, 1, 0, 0, 0, 0]
\nonumber\\
& J_{11}=j[1, 1, 1, 1, 1, 0, 0, 1, 0, 0, 0, 0]
&& J_{12}=j[1, 1, 1, 1, 2, 0, 0, 1, 0, 0, 0, 0]
\nonumber\\
& J_{13}=j[0, 1, 1, 1, 0, 0, 1, 1, 0, 0, 0, 0]
&& J_{14}=j[0, 1, 1, 1, 0, 0, 1, 2, 0, 0, 0, 0] \label{eq:misappc}
\\
& J_{15}=j[0, 1, 1, 1, 0, 0, 2, 1, 0, 0, 0, 0]
&& J_{16}=j[0, 1, 1, 2, 0, 0, 1, 1, 0, 0, 0, 0]
\nonumber\\
& J_{17}=j[0, 1, 2, 1, 0, 0, 1, 1, 0, 0, 0, 0]
&& J_{18}=j[0, 1, 0, 0, 1, 0, 1, 1, 0, 0, 0, 0]
\nonumber\\
& J_{19}=j[0, 1, 0, 0, 1, 0, 1, 2, 0, 0, 0, 0]
&& J_{20}=j[0, 1, 0, 0, 2, 0, 1, 1, 0, 0, 0, 0]
\nonumber\\
& J_{21}=j[0, 1, 0, 1, 1, 0, 1, 1, 0, 0, 0, 0]
&& J_{22}=j[0, 1, 0, 1, 1, 0, 1, 2, 0, 0, 0, 0]
\nonumber\\
& J_{23}=j[0, 1, 0, 1, 2, 0, 1, 1, 0, 0, 0, 0]
&& J_{24}=j[1, 1, 0, 1, 1, 0, 1, 1, 0, 0, 0, 0]
\nonumber\\
& J_{25}=j[0, 1, 1, 1, 1, 0, 1, 1, 0, 0, 0, 0]
&& J_{26}=j[0, 1, 2, 1, 1, 0, 1, 1, 0, 0, 0, 0]
\nonumber\\
& J_{27}=j[0, 2, 1, 1, 1, 0, 1, 1, 0, 0, 0, 0]
&& J_{28}=j[0, 1, 1, 1, 2, 0, 1, 1, 0, 0, 0, 0]
\nonumber\\
& J_{29}=j[0, 1, 1, 2, 1, 0, 1, 1, 0, 0, 0, 0]
&& J_{30}=j[0, 1, 1, 1, 0, 1, 1, 1, 0, 0, 0, 0]
\nonumber\\
& J_{31}=j[1, 1, 0, 0, 1, 1, 1, 1, 0, 0, 0, 0]
&& 
\nonumber
\end{align}


\end{document}